\RequirePackage{fix-cm}
\documentclass[smallextended]{svjour3}       
\smartqed  
\usepackage{graphicx}
\begin{document}

\title{
Fabrication and
Analysis of Three-Layer All-Silicon Interference Optical Filter with
Sub-Wavelength Structure toward High Performance Terahertz Optics
}

\titlerunning{Three-Layer All-Silicon Interference Optical Filter with
Sub-Wavelength Structure }        

\author{
Hironobu~Makitsubo$^{1,2}$
  \and
        Takehiko~Wada$^{2,*}$
  \and
        Hirokazu~Kataza$^{2}$
  \and
        Makoto~Mita$^{2}$
  \and
        Toyoaki~Suzuki$^{3}$
  \and
        Keita~Yamamoto$^{2,4}$
}


\institute{
\and
1:Department of Astronomy, the University of Tokyo, Bunkyo, Tokyo,
113-0033, Japan\\ 
\and
2:Institute of Space and Astronautical Science, Japan Aerospace Exploration Agency,
Sagamihara, Kanagawa 252-5210, Japan\\
\and
3: Graduate School of Science, Nagoya University, Chikusa, Nagoya 464-8602, Japan\\
\and
4:Department of Space and Astronautical Science, 
the Graduate University for Advanced Studies, Sagamihara, Kanagawa 252-5210, Japan\\
\and
*:corresponding author \email{wada@ir.isas.jaxa.jp}
}

\date{Received: 9 Auguest 2016 / Accepted: 7 October 2016}

\maketitle

\begin{abstract}
We propose an all-silicon multi-layer interference filter composed solely of
silicon with sub-wavelength structure (SWS) in order to realize high performance
optical filters operating in the THz frequency region with robustness against
cryogenic thermal cycling and mechanical damage. We demonstrate fabrication
of a three-layer prototype using well-established common
 micro-electro-mechanical systems (MEMS) technologies as a first step
 toward developing practical filters. The measured 
transmittance of the three-layer filter agrees well with the theoretical
 transmittances
calculated by a simple thin-film calculation with effective refractive indices as well as a
rigorous coupled-wave analysis simulation. We experimentally show that
 SWS
layers can work as homogeneous thin-film interference layers with effective refractive
indices even if there are multiple SWS layers in a filter.

\end{abstract}

\section{Introduction}
\label{intro}
Optical band-pass filters (OBPFs) are essential components for
photometric observations to obtain both color and spatial information
simultaneously. High performance OBPFs with rectangular transmission
spectrum profiles are critical for precision and dependability. In the
THz frequency region, robustness against cryogenic thermal cycling is
required to obtain high-sensitivity observations because the OBPFs must
be cooled to avoid thermal background noise from the filter
itself. Mechanical strength is also required in space applications for
components to endure stresses from rocket launching. Furthermore,
robustness in filters against rapid decompression and extremely large
acoustic vibrations is required for the next-generation warm-launch
cryogenic astronomical telescopes, such as SPICA \cite{nakagawa12}, as there will be
no vacuum chambers to protect OBPFs.

It is, however, still difficult by conventional methods to develop high
performance OBPFs for the THz region that have the necessary robustness
against cryogenic thermal cycling and mechanical damage. Thin-film
multi-layer interference filters are widely used in optical and near
infrared wavelength regions \cite{Quijada04}. Thin-film filters enable us to realize
high performance OBPFs; by increasing the number of layers, nearly
rectangular-shaped profiles in the transmission spectra can be obtained
\cite{Macleod10}. It is favorable for applying the thin-film filters to the THz region,
but there is difficultly that only a few materials are known to have
good transparency and physical robustness in the region. Furthermore,
the conventional thin-film
filters are coatings of hetero-materials with different coefficients of
thermal expansion (CTE), and thus these heterogeneous multi-layer
structures inherently suffer large stresses in cryogenic thermal
cycling.  Table~\ref{tab:CTE} shows the CTE of PbTe and CdSe, which is
commonly used for thin-film filters at the THz region \cite{seeley81}. 
Metal mesh filters are also commonly used
as 10 THz-region OBPFs \cite{Sako08}, but designing those with
transmission spectra
with near-rectangular profiles and broad pass-band is difficult in
principle, and furthermore freestanding metal thin-films are
mechanically fragile.

\begin{table}
\caption{CTE of materials used for 10\,THz filters at 300\,K}
\label{tab:CTE}       
\begin{tabular}{lll}
\hline\noalign{\smallskip}
       & PbTe & CdTe  \\
\noalign{\smallskip}\hline\noalign{\smallskip}
CTE & $1.98\times10^{-5}$ (poly \cite{nii64}) & $4.13\times10^{-6}$ (a
	 direction \cite{iwanaga00}) \\
($\mathrm{K}^{-1}$)    & $2.04\times10^{-5}$ (crystal \cite{houston68}) &
	 $2.76\times10^{-6}$ (c direction \cite{iwanaga00}) \\
\noalign{\smallskip}\hline
\end{tabular}
\\
\end{table}

\begin{figure}
  \includegraphics[width=1.0\textwidth]{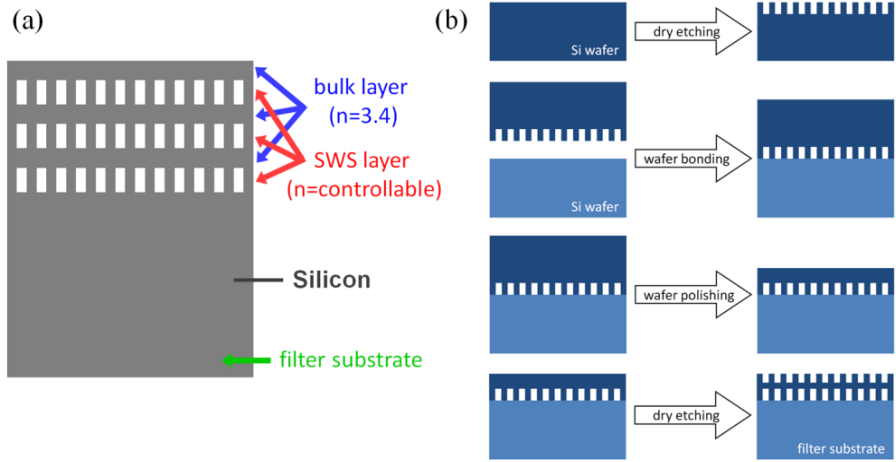}
\caption{
Cross-sectional schematic of (a) all-silicon multi-layer interference
 filter with SWS and (b) fabrication procedure of a three-layer
 all-silicon filter with SWS using MEMS technologies.
}
\label{fig:1}       
\end{figure}

To overcome these difficulties, we proposed an all-silicon
(mono-material) multi-layer interference filter with sub-wavelength
structure (SWS) \cite{Wada10}; Fig.~\ref{fig:1} (a) shows a cross-sectional schematic of
the filter. All-silicon filters have superior robustness against
cryogenic thermal cycling because it is free from CTE mismatching that
plagues conventional heterogeneous thin-film filters. Furthermore,
all-silicon filters have high throughput in the THz region owing to low
absorption of high-resistivity silicon \cite{Edwards85,Dai04}. Silicon is also
well-suited for mono-material filters because not only it displays good
physical and chemical stability but also micro-fabrication processes
have been well established for it. We introduce SWS layers to control
the refractive indices of each interference layer. According to the
effective medium theory (EMT) \cite{Brauer94}, a SWS layer behaves as a homogenous
medium layer with an effective refractive index. Recently, SWS surfaces were
being used widely for optical elements following the fast developments in
micro-fabrication technologies \cite{Kikuta03}. Multi-layer SWS surfaced
devices was also proposed and developed, such as the three-layer
birefringent quarter-wave plate consisting of two materials \cite{Yu06}.

In this paper, we report on the fabrication of a three-layer all-silicon
filter with SWS to demonstrate our proposed fabrication method. We also
compare the spectral performance with theoretical calculations to
verify that the SWS layers work as homogeneous thin-film interference
layers even if there are multiple SWS layers in a filter. The
three-layer filter is a first step for future practical multi-layer
OBPFs, as in principle multi-layer structures can be fabricated to any
degree by iterating the three-layer fabrication cycle.

%


\section{Fabrication}
\label{fab}
Fig.~\ref{fig:1} (b) shows the fabrication procedure of a three-layer
all-silicon filter with SWS. We used well-established common
micro-electro-mechanical systems (MEMS) technologies: photolithography
with dry etching, wafer direct bonding, and wafer chemical mechanical
polishing (CMP). We used non-doped high-resistivity ($>10\,\mathrm{k\Omega \mbox{-}cm}$)
single-crystal silicon (100) wafers fabricated using the floating zone
(FZ) method. The diameter and thickness of the wafer was 50\,mm and
0.5\,mm, 
respectively. Both sides of the wafer were polished to a mirror
surface by CMP. Our AFM observations of the surface after the
CMP showed roughness of 0.4\,nm RMS.
 At first, we performed photolithography and dry etching
to fabricate a SWS layer on the surface of the wafer. We used deep
inductively coupled plasma reactive ion etching (ICP-RIE) \cite{Marty05} equipment
(Sumitomo Precision Products MUC-21) with SF$_6$ as etching gas
and CF$_4$ as passivation, 
and fabricated deep holes by the Bosch process \cite{Laermer96}. 
Next, we applied wafer
direct bonding to create a second layer. The wafer with SWS and a bulk
wafer were bonded by a surface activated wafer bonding (SAB) technique
\cite{Takagi96} using a room-temperature wafer bonding machine (Mitsubishi Heavy
Industries MWB-04R). Then, we thinned the wafer to control the thickness
of the bulk layer using a grinder (Musashino-denshi MA-200) and diamond
particles as abrasives. The wafer was finally polished by CMP to obtain
a mirror surface for future SAB process. We further performed
photolithography and ICP-RIE processing to fabricate an additional SWS
layer. Thus, a three-layer all-silicon filter with SWS was produced.

\begin{figure}
  \includegraphics[width=1.0\textwidth]{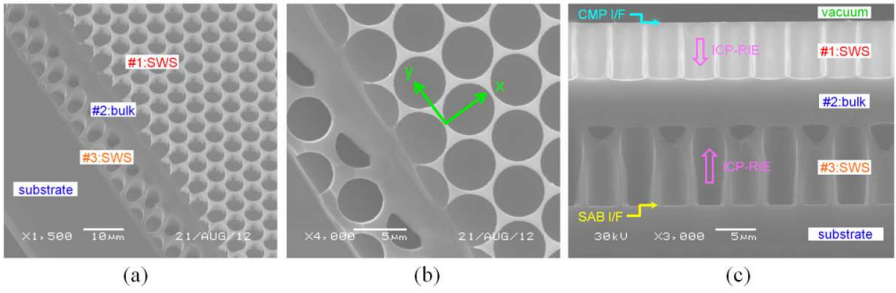}
\caption{
SEM images of the fabricated three-layer all-silicon filter with SWS: (a) oblique view, (b) top view, and (c) cross-sectional view.
}
\label{fig:2}       
\end{figure}

Fig.~\ref{fig:2} shows scanning-electron microscope (SEM, Japan Electron Optics
Laboratory JSM-5510) images of the fabricated three-layer all-silicon
filter with SWS. We measured SWS dimensions and thicknesses of each
layer from the SEM images. For convenience, each layer is designated a
symbol \#(layer number):(medium structure); that is, the SWS layer on
the vacuum side is \#1:SWS, the bulk layer is \#2:bulk, and the SWS
layer on the substrate side is \#3:SWS. The hole diameters of layers
\#1:SWS and \#3:SWS are 4.9\,$\mu$m and 4.3\,$\mu$m, respectively. The distances
between centers of two adjacent holes in layers \#1:SWS and \#3:SWS are
5.2\,$\mu$m. Thus, the layer porosities of \#1:SWS and \#3:SWS are 80\% and
62\%, respectively. The layer thicknesses of \#1:SWS, \#2:bulk, and
\#3:SWS are 6.8\,$\mu$m, 6.0\,$\mu$m, and 9.8\,$\mu$m, respectively. We note that
the interface between layers \#1:SWS and \#2:bulk has a round shape
resulting from the ICP-RIE process, so that we define the thickness of
the \#1:SWS layer as a mean depth in the \#1:SWS cylindrical holes. The
thickness of the \#3:SWS layer is defined similarly. We also note that
both holes of layers \#1:SWS and \#3:SWS can be simultaneously observed
even in the top view image, Fig.~\ref{fig:2} (b), because the three-layer filter
was cleaved at a slant.

\section{Theoretical calculation}
\label{sec:theory}
\subsection{Methods}
\label{subsec:method}
We calculated a theoretical transmittance of the fabricated filter
(three interference layers on one side of the filter substrate) by
following procedure. First, we calculated transmittances and
reflectances of only the three interference layers using two calculation
methods: a three-layer full simulation using rigorous coupled-wave
analysis (RCWA) \cite{Moharam86} and a simple three-layer thin-film calculation with
effective refractive indices, in order to verify theoretically that SWS
layers can work as homogeneous thin-film layers even in filters with
multiple SWS layers. Finally, by considering multiple reflections in the
filter substrate, we calculated transmittances of the entire filter.

In the three-layer full RCWA simulation, we made a calculation model
using the measured SWS dimensions and thicknesses obtained from the SEM
images. We also took into account the distance between the central axes
of the \#1:SWS holes and the \#3:SWS holes. 
In the simulation, the incident angle
was $0^{\circ}$ and the order of the Fourier coefficients was between -10th and
+10th for both x and y directions, marked in Fig.~\ref{fig:2} (b). We calculated
unpolarized transmittances by averaging the two orthogonal linear
polarizations. The RCWA simulation was performed by commercial software
(RSoft Design Group DiffractMOD).

\begin{figure}
  \includegraphics[width=0.9\textwidth]{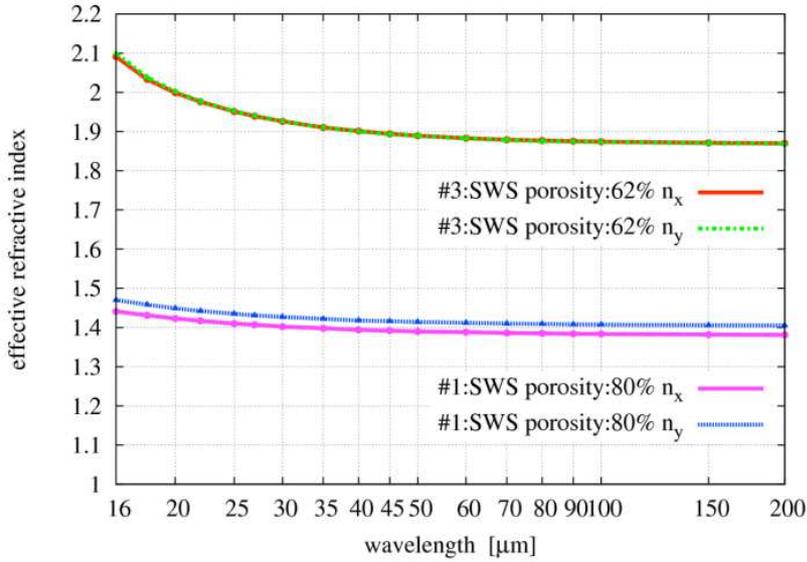}
\caption{
Effective refractive indices of layers \#1:SWS and \#3:SWS for each
 orthogonal linear polarization as a function of wavelength; here,
 $\mathrm{n_x}$ and $\mathrm{n_y}$ are the respective refractive indices
 for linear polarization, for electric fields parallel to the x and y
 directions in Fig~\ref{fig:2}(b).
 Note X-axis is in logarithmic scale.
}
\label{fig:3}       
\end{figure}

For the simple three-layer thin-film calculation, we estimated the
effective refractive indices of layers \#1:SWS and \#3:SWS by the
following procedure. First, we calculated a reflectance for a given
light wavelength of one SWS-layer filter (only one SWS-layer on a
substrate) as a function of the SWS-layer thickness. This calculation
itself was performed using the RCWA simulation. Then, we found the
thicknesses where the reflectance takes local minimums. These
thicknesses must be odd times of the quarter-wave optical thickness
because of interference conditions, because we assumed that the SWS layer
was a homogenous dielectric layer. Thus, we determined the effective
refractive index $n_{\mathrm{eff}}$ from formula, $n_{\mathrm{eff}} = \lambda/4d_{\lambda/4}$, where $\lambda$ was the
light wavelength in vacuum and $d_{\lambda/4}$ was the smallest thickness that
gives the local reflectance minimum. Fig.~\ref{fig:3} shows the effective
refractive indices of layers \#1:SWS and \#3:SWS for each orthogonal
linear polarization as a function of wavelength. The effective
refractive indices exhibit wavelength dispersion, signifying that
zeroth-order EMT \cite{Brauer94} is no longer valid at shorter wavelength
regions. Diffraction, however, does not occurred in the wavelength
regions ($\lambda = 16\mbox{--}200\,\mu\mathrm{m}$). Hence, with these effective refractive indices,
we can calculate transmittances of the three interference layers for
each orthogonal linear polarization, and then by averaging the two, we
can calculate the unpolarized transmittances.

In the above calculations, the refractive index of silicon was set to
3.42 (no wavelength dispersion). In the three-layer calculation, we did
not consider absorption in silicon because the three interference layers
were sufficiently thin that absorption is negligible. In the substrate
calculation, however, we considered the extinction coefficients of
silicon \cite{Edwards85,Dai04} because the substrate was thick enough
that absorption could not be ignored. We note that the measured transmission spectrum of the
bulk silicon wafer agreed well with the calculation using the
above-mentioned optical constants.

\subsection{Results}
\label{subsec:results}
\begin{figure}
  \includegraphics[width=0.9\textwidth]{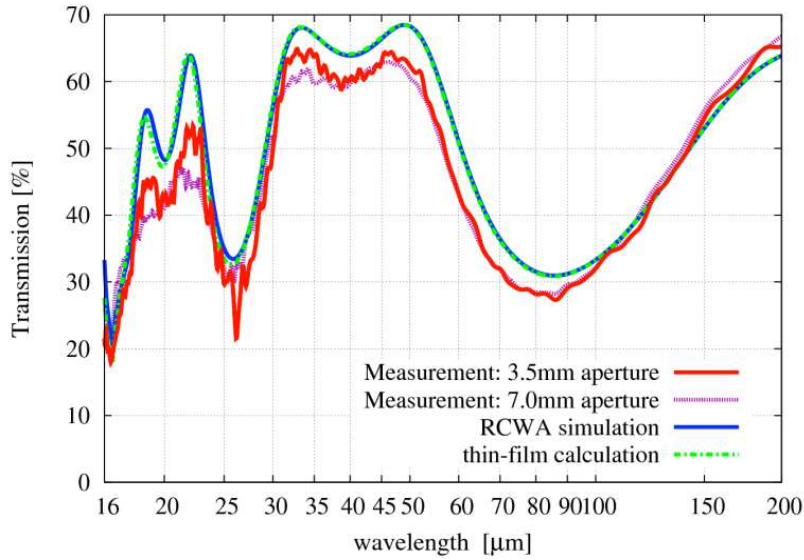}
\caption{
Measured and calculated transmission spectra of the three-layer
 all-silicon filter with SWS.
 Note X-axis is in logarithmic scale.
}
\label{fig:4}       
\end{figure}

Fig.~\ref{fig:4} shows theoretical transmission spectra of the three-layer
filter calculated by a three-layer full RCWA simulation and by a simple
three-layer thin-film calculation with the effective refractive indices
shown in Fig.~\ref{fig:3}. The calculated transmittance from the full RCWA
simulation agrees well with that from the simple thin-film
calculation. Therefore, theoretically, the SWS layers can function as
homogeneous thin-film layers even if there are multiple SWS layers in a
filter.

We remark that we calculated various transmittances using the
three-layer full RCWA simulation by changing the distance between the
central axes of the \#1:SWS and \#3:SWS holes, and that the results
overlap in the vertical axis resolution in Fig.~\ref{fig:4}. 
Thus, we do not need
to align the centers of each SWS-layer hole during wafer bonding,
enabling us to produce multi-layer all-silicon filters by current MEMS
technologies.

\section{Experimental results}
\label{sec:exp-result}
We measured the transmittance of the three-layer filter by a Fourier
transform infrared (FT-IR) spectrometer (BOMEM DA8) at room temperature
and under vacuum conditions ($<$ 0.3\,Torr). A mercury lamp, a Mylar film,
and a deuterated triglycine sulfate (DTGS) pyroelectric detector were
used as light source, beam-splitter for the Michelson interferometer,
and light detector, respectively. The aperture sizes of the light beam
were 3.5\,mm and 7.0\,mm; the beam incident angle was $0^{\circ}$.
The measured
transmission spectra are presented in Fig.~\ref{fig:4}. The spectral resolution
was 4.0\,cm$^{-1}$ and the transmittance accuracy was about 10\% as evaluated
from the scattering of the measured values taken on different days. The
fluctuation of the measured transmittance around $\lambda=26\,\mu\mbox{m}$ is due to the
degradation in measurement accuracy resulting from poor signal-to-noise
ratio.
We performed multiple thermal cycling down to LN$_{2}$ temperature
and confirmed that the transmission spectra did not change 
in our measurement accruracy.

As seen in Fig.~\ref{fig:4}, measured and theoretical transmittances agree
well. In particular, the measured wavelengths where transmittances take
local maximum or minimum values correspond well to theoretical
values. This is the experimental confirmation of the functioning of the
SWS layers as homogeneous thin-film interference layers. Thus, we can
save time when designing practical extended multi-layer all-silicon
filters because we can estimate the transmission spectra using only the
simple thin-film calculation instead of the time-consuming multi-layer
full RCWA simulation.

\section{Discussion}
\label{sec:discussion}
All-silicon filters can achieve far superior throughput in transmission
pass-band because the absorption of silicon in the THz region is
sufficiently small. In the fabricated three-layer filter, however, the
theoretical maximum transmittance is 70\% because of Fresnel reflection
at the backside interface between substrate and vacuum. The loss of
transmittance by Fresnel reflection vanishes when we fabricate
interference layers on both sides of the filter substrate; the maximum
transmittance reaches 100\% with the precise design of the interference
layers. 

\begin{figure}
  \includegraphics[width=0.50\textwidth]{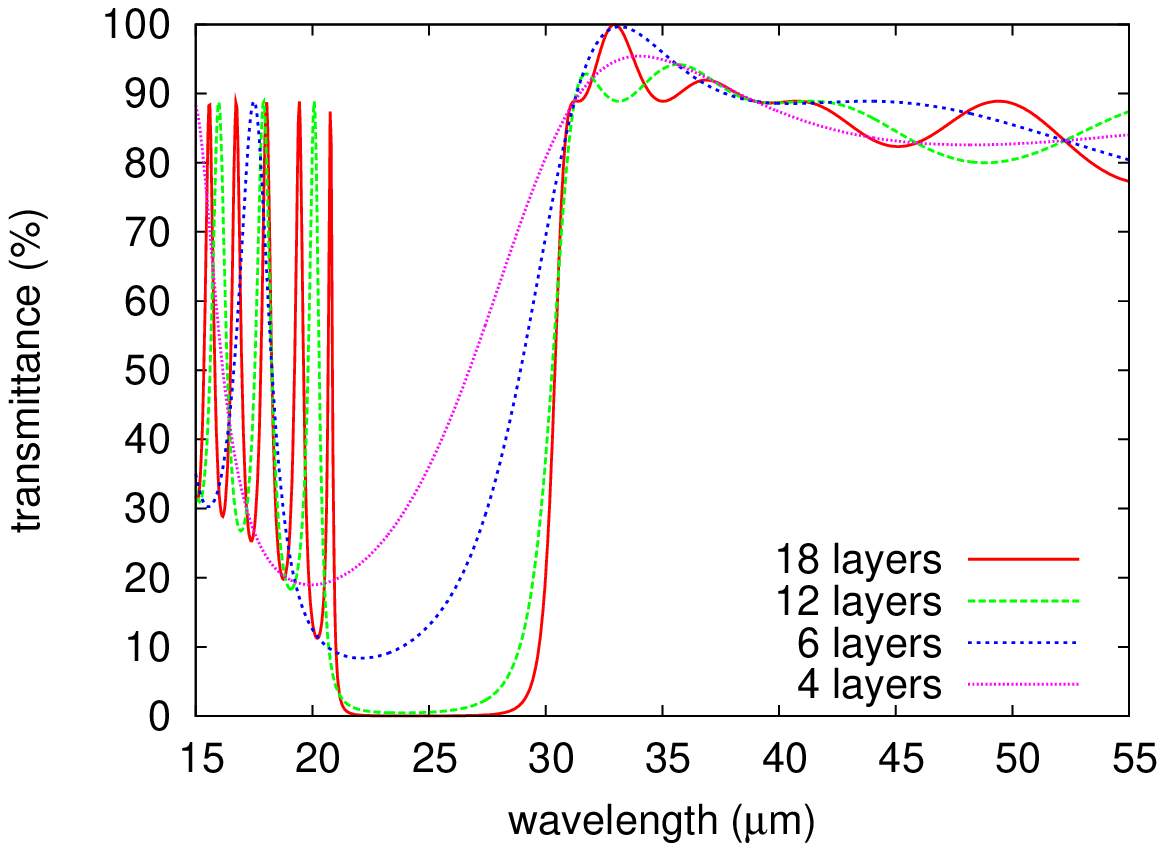}
  \includegraphics[width=0.50\textwidth]{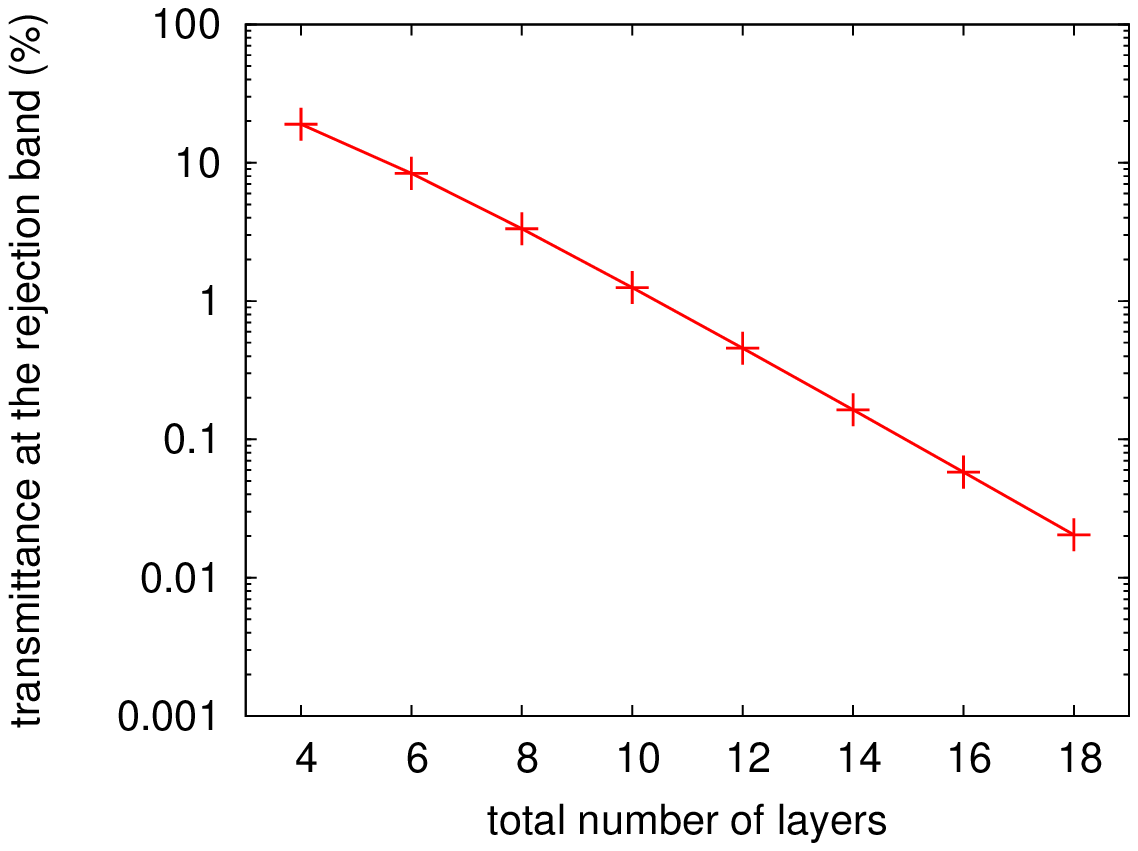}
\caption{
(left) Calculated transmission spectra of LWPFs with a QWOT structure of 
$[\mathrm{S}\,|\,0.5L\,(1.0H\,1.0L)^m\,0.5H\,|\,\mathrm{V}]$,
where $H$ and $L$ are the QWOT of bulk Si ($n=3.4$) and SWS ($n_{\mathrm{eff}}=2.0$) layers.
In the calculation, we omitted the Fresnel reflection
at the backside surface of the substrate, where a SWPF 
is commonly fabricated to configure a band-pass filter.
(right) Transmittance at the bottom of the rejection band 
 as a function of the total number of layers ($2 + 2 \times m$). 
\label{fig:5}       
}
\end{figure}

The rejection-band transmittance of the three-layer filter is
about 30\%, which is not sufficient for practical use (typically
below 0.1\%). 
In order to estimate the number of layers which is necessary for practical use,
we have designed long wave pass filters (LWPFs) which are commonly 
used to configure a band-pass filter together 
with a short wave pass filter (SWPF) at the other side of the substrate.
We have designed the LWPFs with quarter wave optical thickness (QWOT)
structure of 
$[\mathrm{S}\,|\,0.5L\,(1.0H\,1.0L)^m\,0.5H\,|\,\mathrm{V}]$, 
where
S and V represent Si substrate ($n=3.4$) and the vacuum, 
and $H$ and $L$ are the QWOT of Si bulk ($n=3.4$) and SWS ($n_{\mathrm{eff}}=2.0$) layers, respectively.
The total number of layer is $2 + 2 \times m$.
In a sixteen-layer filter (m=7), the transmittance of the rejection-band falls
below 0.1\% and the transmission spectrum becomes more rectangular in
shape as the number of layers increases (Fig.~\ref{fig:5}).

The flatness of each interference layer is the key in achieving high
performance OBPFs. Different thicknesses of an interference layer within
a filter result in different transmission spectra at various positions
over the filter. Therefore, especially steep shapes in a transmission
spectrum are broken because an observed transmission spectrum from an
aperture is an integration of transmission spectra at various filter
positions within the aperture. From a filter designing perspective, each
layer thickness should be flat to within $\pm 0.1\,\mu\mathrm{m}$ for an OBPF around
the $30\,\mu\mathrm{m}$ wavelength region. This requirement decreases at longer
wavelength to $\pm 1\,\mu\mathrm{m}$ for the $300\,\mu\mathrm{m}$ wavelength region.

In our fabricated three-layer filter, the layer thicknesses of \#1:SWS
and \#3:SWS are uniform at various filter positions within the
measurement accuracy ($\pm 0.05\,\mu\mathrm{m}$). The layer thickness of \#2:bulk is,
however, not uniform; at the transmission-measured position the
thickness is $6.0\,\mu\mathrm{m}$, but the thicknesses at +5 mm, -5 mm, and -10 mm
away from the transmission-measured position are $6.7\,\mu\mathrm{m}$, $5.0\,\mu\mathrm{m}$, and
$3.5\,\mu\mathrm{m}$, respectively. Clearly, the \#2:bulk layer has a large gradient
in thickness, which is not adequate in practical applications. Indeed,
the measured transmittance from the 7.0 mm aperture beam is clearly
lower than that from the 3.5 mm aperture beam for wavelengths around 
$\lambda=19\,\mu\mathrm{m}$ and $\lambda=22\,\mu\mathrm{m}$
 where the transmission spectrum rise steeply (see
Fig.~\ref{fig:4}). The gradient of the \#2:bulk layer thickness is also considered
as a reason for disagreement between the measured and calculated
transmittances around $\lambda=19\,\mu\mathrm{m}$ and
$\lambda=22\,\mu\mathrm{m}$. 
This gradient may come
from a $\mu$m-scale particle between the filter and the holder of the
grinder during CMP. To produce a flat bulk silicon layer for practical
applications, we are planning to use a device layer of a
silicon-on-insulator (SOI) wafer.

\section{Conclusion}
\label{sec:conclusion}
We have fabricated a three-layer all-silicon interference filter with
SWS by combining well-established common MEMS technologies. The measured
transmission spectrum of the three-layer filter agreed well with
theoretical calculations using not only a three-layer full RCWA
simulation but also a simple three-layer thin-film calculation with
effective refractive indices determined by a one-layer RCWA
simulation. We conclude that the SWS layers function as homogeneous
thin-film interference layers even if there are multiple SWS layers in a
filter, and that we can design transmission spectra of practical
extended multi-layer all-silicon filters with SWS using a simple
thin-film calculation with effective refractive indices.


\begin{acknowledgements}
The authors are deeply grateful to Mr. Jun Utsumi and Mr. Masahiro Kato
 at Mitsubishi Heavy Industries, Ltd. for supporting the
 room-temperature wafer bonding process. This study was supported by the
 FY2009 and FY2012 Japan Aerospace Exploration Agency President Fund,
 and FY2010 ISAS/JAXA Basic R\&D on onboard equipment.
\end{acknowledgements}




\end{document}